\documentclass[sigconf]{acmart}
\usepackage[T1]{fontenc}
\usepackage{subfigure,epsfig}
\usepackage{algorithm}
\usepackage{algorithmic}
\usepackage{multirow}
\usepackage{float}
\usepackage{url}
\usepackage{bm}
\usepackage{enumitem}

\usepackage{array}
\usepackage{booktabs}
\renewcommand\arraystretch{0.9}
\setlist[itemize]{leftmargin=*}
\setlist[enumerate]{leftmargin=*}

\newcommand{\ie}{\emph{i.e., }}
\newcommand{\eg}{\emph{e.g., }}

\newcommand{\wrt}{\emph{w.r.t. }}

\def\BibTeX{{\rm B\kern-.05em{\sc i\kern-.025em b}\kern-.08emT\kern-.1667em\lower.7ex\hbox{E}\kern-.125emX}}
    
\copyrightyear{2020}
\acmYear{2020}
\setcopyright{acmcopyright}\acmConference[MM '20]{Proceedings of the 28th ACM International Conference on Multimedia}{October 12--16, 2020}{Seattle, WA, USA}
\acmBooktitle{Proceedings of the 28th ACM International Conference on Multimedia (MM '20), October 12--16, 2020, Seattle, WA, USA}
\acmPrice{15.00}
\acmDOI{10.1145/3394171.3413556}
\acmISBN{978-1-4503-7988-5/20/10}
\begin{document}
\fancyhead{}

\title{Graph-Refined Convolutional Network for Multimedia Recommendation with Implicit Feedback}
\author{Yinwei Wei}
\affiliation{
    \institution{Shandong University}
}
\email{weiyinwei@hotmail.com}

\author{Xiang Wang$^\S$}
% \authornote{Xiang Wang and Liqiang Nie are the corresponding authors.}
\affiliation{
    \institution{National University of Singapore}
}
\email{xiangwang@u.nus.edu}

\author[a]{Liqiang Nie$^\S$}
% \authornote{}%Co-corresponding author}
\affiliation{
    \institution{Shandong University}
}
\email{nieliqiang@gmail.com}

\author{Xiangnan He}
\affiliation{
    \institution{University of Science and Technology of China}
}
\email{xiangnanhe@gmail.com}

\author{Tat-Seng Chua}
\affiliation{
    \institution{National University of Singapore}
}
\email{chuats@comp.nus.edu.sg}
\thanks{$^\S$Xiang Wang and Liqiang Nie are the corresponding authors.}
\begin{abstract}

Reorganizing implicit feedback of users as a user-item interaction graph facilitates the applications of graph convolutional networks (GCNs) in recommendation tasks.
In the interaction graph, edges between user and item nodes function as the main element of GCNs to perform information propagation and generate informative representations.
Nevertheless, an underlying challenge lies in the quality of interaction graph, since observed interactions with less-interested items occur in implicit feedback (say, a user views micro-videos accidentally).
This means that the neighborhoods involved with such false-positive edges will be influenced negatively and the signal on user preference can be severely contaminated.
However, existing GCN-based recommender models leave such challenge under-explored, resulting in suboptimal representations and performance.

In this work, we focus on adaptively refining the structure of interaction graph to discover and prune potential false-positive edges.
Towards this end, we devise a new GCN-based recommender model, \emph{Graph-Refined Convolutional Network} (GRCN), which adjusts the structure of interaction graph adaptively based on status of model training, instead of remaining the fixed structure.
In particular, a graph refining layer is designed to identify the noisy edges with the high confidence of being false-positive interactions, and consequently prune them in a soft manner.
We then apply a graph convolutional layer on the refined graph to distill informative signals on user preference.
Through extensive experiments on three datasets for micro-video recommendation, we validate the rationality and effectiveness of our GRCN.
% over several state-of-the-art models.
Further in-depth analysis presents how the refined graph benefits the GCN-based recommender model.

\end{abstract}

\vspace{-5pt}

\maketitle

% \newpage
\section{Introduction}
With the high prevalence of the Internet, people have access to large amounts of online multimedia content, such as movies, news, and music.
In multimedia content sharing platforms (\eg Instagram, YouTube, and Tiktok), multimedia recommendation has been a core service to help users identify items of interest. 
At the core of the recommendation is exploiting multimedia contents of items and historical behaviors of users (\eg views, clicks) to capture user preference and consequently suggest a ranking list of items.

Learning informative representations of users and items has become a central theme in multimedia recommender systems.
Early works like VBPR~\cite{VBPR} and ACF~\cite{ACF} integrate multimedia contents (\eg visual features) and ID embeddings of items together in the traditional collaborative filtering (CF) framework.
However, these models limit to explore underlying relationships among users and items, since only direct user-item interactions are taken into consideration.
More recently, inspired by the success of graph convolutional networks~(GCNs)~\cite{MM1,MM2,MM3,MM4}, some efforts~\cite{PinSAGE,NGCF,KGAT,MMGCN,MGAT} have been made to organize user behaviors as a bipartite user-item graph and integrate multi-hop neighbors into representations.
Such GCN-based recommender models benefit from powerful representation ability of GCN and have achieved the state-of-the-art performance.

Despite their remarkable performance, we argue that the fixed interaction graphs built upon implicit feedback are highly likely to contain noisy behaviors of users.
For example, a user might click some videos shared by her/his friends or even accidentally, while she has no interest in these videos.
Such false-positive behaviors appear in the interaction graph as edges between user and item nodes, which are treated equally with the true-positive interactions.
When performing information propagation of GCNs, the neighborhoods around these false-positive edges will be influenced negatively and the signals on user preference can be severely contaminated.
This is consistent to the vulnerability of GCNs against structure perturbations~\cite{RAAA,Blackbox}.
Therefore, we further argue that the performance of GCN-based recommendation can be significantly degraded by adding a few edges of false-positive interactions in the graph.

\begin{figure}
% 	\centering
    \includegraphics[width=0.4\textwidth]{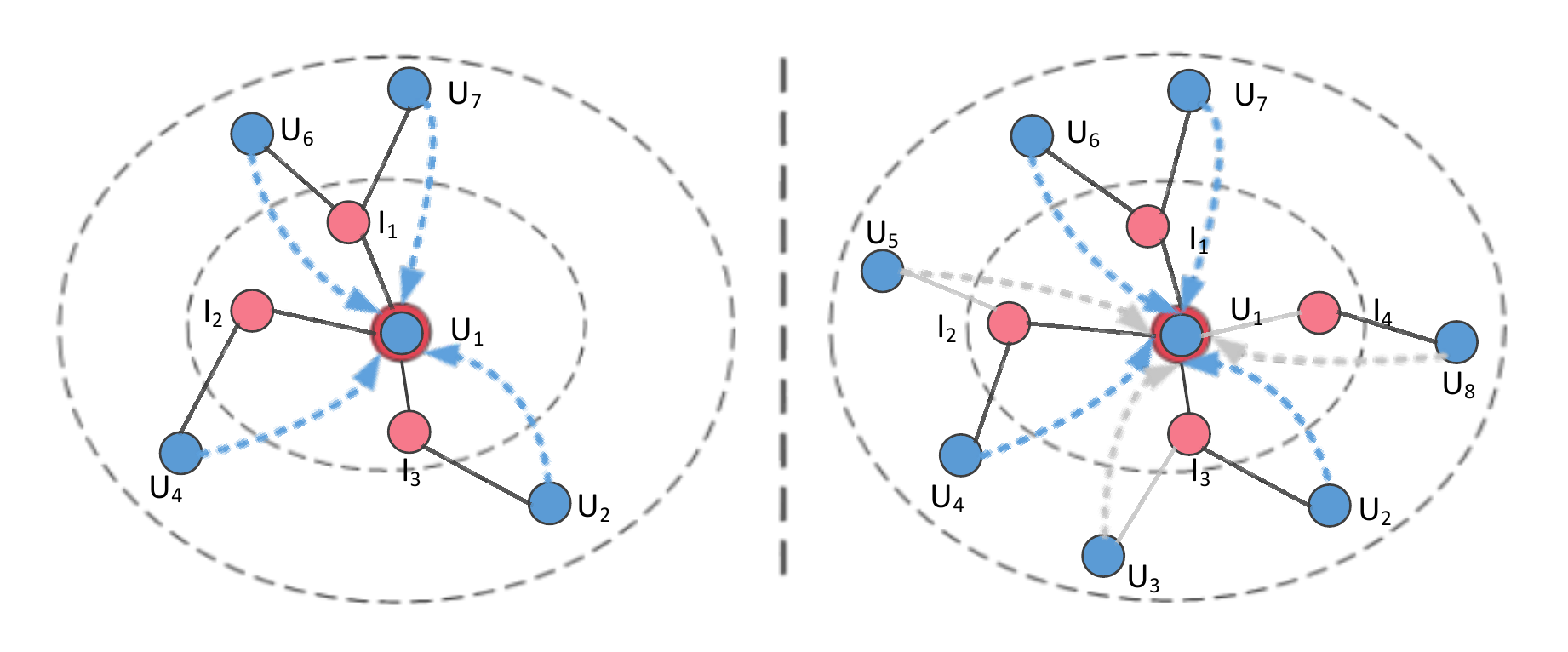}
    % \vspace{-10pt}
    \caption{Illustration of information flows (denoted as dashed curve) caused by true-positive and false-positive interactions in the GCN-based recommender models.}
    \vspace{-10pt}
	\label{fig:intro}
\end{figure}

\vspace{5pt}
\noindent\textbf{Running Example.} Figure~\ref{fig:intro} illustrates how the false-positive feedback disturbs the information propagation of the GCN-based recommender model.
In the left subfigure, the clean graph involves user-item connections, each of which indicates that a user is truly interested in the item (\ie solid black lines like $(U_{4},I_{2})$);
whereas, the graph in the right side additionally includes user interactions with less interested items (\ie solid grey lines like $(U_{5},I_{2})$).
Along with the graph structures, GNN-based recommenders conduct the information propagation mechanism~\cite{GCN,GAT,GraphSAGE} to distill collaborative signal and generate collaborative embeddings of users and items.
However, due to the structure difference, not only signal pertinent to user preference (\ie dashed blue curves like $(U_{4},U_{1})$), but also noisy information (\ie dashed grey curves like $(U_{5},U_{1})$) are aggregated into $U_{1}$'s collaborative embeddings.
Further stacking more graph convolutional layers will introduce more noisy signal from multi-hop neighbors.
As a result, the representations can be contaminated, and the performance of GNN-based recommenders can be severely degraded.

\vspace{5pt}\noindent\textbf{Present Work.}
To tackle this challenge, we aim to identify and prune the edges that are potential false-positive interactions.
Towards this end, we develop a new model, \textit{Graph-Refined Convolutional Network}~(GRCN), which exploits the rich content of items and historical behaviors of users to adaptively refine the structure of interaction graph.
It consists of three components: graph refining, graph convolutional, and prediction layers.
Specifically, the graph refining layer hires the neighbor routing mechanism~\cite{DisenGCN} to refine a prototypical network, highlighting the user preference towards a item prototype \wrt content in individual modalities.
Intuitively, for a given user, an affinity between the target item and her/his prototype reflects the confidence of the target item being true positive in each modality.
Then a pruning operation is adopted to prune the edges according to their affinity scores, to corrupt the propagation of their noisy signal. 
Built upon the refined interaction graph, we apply a graph convolutional layer to obtain the high-quality collaborative embeddings of users and items.
Finally, in the prediction layer, we predict how likely a user adopts an item by calculating the similarity between their representations. 
To demonstrate our proposed method, we conduct extensive experiments on three public datasets.
Empirical results validate that our proposed model outperforms the state-of-the-art baselines like MMGCN~\cite{MMGCN}, DisenGCN~\cite{DisenGCN}, and GAT~\cite{GAT}. 
Moreover, the visualization of the learned user and item embeddings offers a reasonable explanations on why the graph refining operations boosts the GCN-based recommendation method.  
In a nutshell, our contributions are summarized as follows:
\begin{enumerate}
    \item We explore the influence of implicit feedback to the GCN-based recommendation model. To the best of our knowledge, this is the first attempt to solve the implicit feedback problem against the high-order connectivity. 
    \item  We develop a new method GRCN, which adaptively refines the structure of user-item interaction graph to harness the applications of GNNs in recommendation tasks.
    \item Extensive experiments in three real-world datasets validate the rationality of our assumptions and the effectiveness of our method. Our codes are available in \url{https://github.com/weiyinwei/GRCN}.
\end{enumerate}
\section{methodology}
\subsection{Preliminary}
Suppose there are numbers of historical interaction records~(\textit{i.e.} implicit feedback) between users and items. We collect a set $\mathcal{U}$ of $N$ users and a set $\mathcal{I}$ of $M$ items from the records. 
Beyond the interaction signal, the multimodal features of items are extracted from their content involving the visual, acoustic, and textual modalities, which are denoted as $v$, $a$, and $t$, respectively. 
For a item $i\in\mathcal{I}$, we denote its feature vector as $\mathbf{i}_m\in\mathcal{R}^{M\times D_m}$, where $m\in\mathcal{M}=\{v, a, t\}$ is the indicator of multiple modalities and $D_m$ is the dimension of the vector. 

To conduct the graph convolutional operations, we construct a user-item interaction graph $\mathcal{G}=\{\mathbf{E}, \mathbf{A}\}$, which follows the GCN-based recommendation~\cite{NGCF, MMGCN}. 
In particular, $\mathbf{E}\in \mathcal{R}^{D\times(N+M)}$ denotes the trainable embedding matrix of nodes~(\textit{i.e.} users and items), where $D$ represents the dimension of the embedding. 
And, $\mathbf{A}\in \mathcal{R}^{N\times M}$ is the symmetric matrix reflecting the connections of user and item pairs. 
Given a user $u\in \mathcal{U}$ and a item $i\in \mathcal{I}$, we denote $\mathbf{A}_{u,i}=1$ if $u$ has interacted with $i$; otherwise, $\mathbf{A}_{u,i}=0$. 
   
\subsection{Model Framework}
\begin{figure*}
	\centering
    \includegraphics[width=1\textwidth]{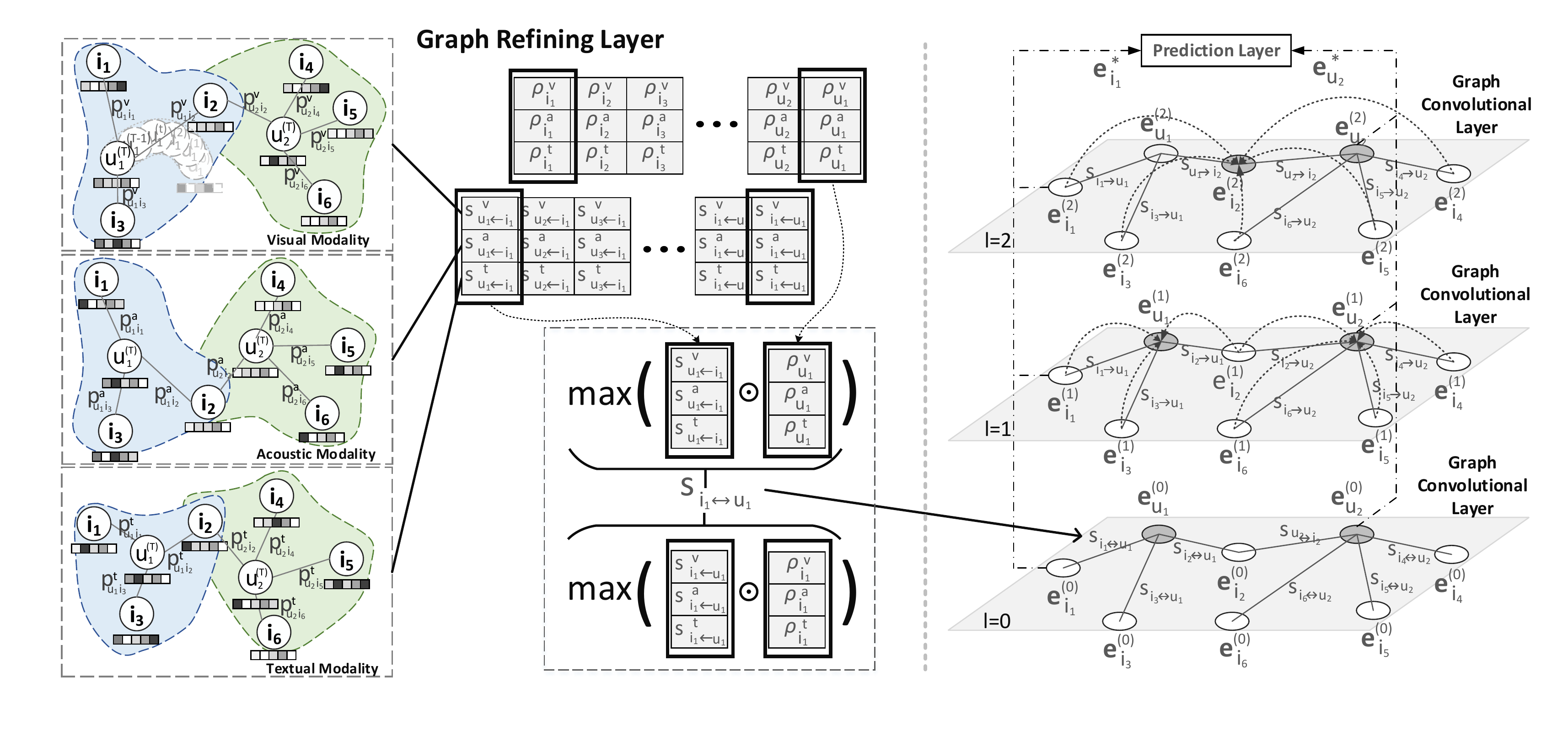}
	  \vspace{-20pt}
    \caption{Schematic illustration of our proposed model. It consists of three components, namely graph refining layer, graph convolutional layer, and prediction layer.}
	\label{fig:method}
	    \vspace{-5pt}
\end{figure*}

In this section, we detail our proposed model. As illustrated in Figure~\ref{fig:method}, the model consists of three components: 1) the graph refining layer that adjusts the graph structure by identifying and pruning the noisy edges in interaction graph; 2) the graph convolutional layer which performs the graph convolutional operations on the refined graph to enrich the embeddings of items and users; and 3) the prediction layer that infers the interaction of each user and item pair. 
\subsubsection{Graph Refining Layer}
To refine the structure of constructed interaction graph, we work under the reasonable assumption that the content of item belonging to false-positive interaction is far from the user preference. 
Therefore, we introduce the prototypical network to learn user preference to the content information, and then prune the noisy edges 
according to the confidence of edges being the false-positive interactions.  

\vspace{3pt}\par\noindent\textbf{Prototypical Network}. 
Intuitively, each user preference could be learned from the content of items which directly connect to the user node in the user-item graph. 
However, since there are some noisy edges in the graph, it is hard to immediately model the user preference with the neighbor nodes. 
Inspired by the idea of prototype learning~\cite{PrototypeNetwork}, we regard the user preference as her/his prototype in a metric space and harness a prototypical network to approach it. 

For this goal, the content signal of item is projected into a metric space to distill the informative features related to the user preference, as
	\begin{equation}
		\bar{\mathbf{i}}_m = leaky\_relu(\mathbf{W}_m\mathbf{i}_m+\mathbf{b}_m)
	\end{equation}
where $leaky\_relu(\cdot)$, $\mathbf{W}_m\in\mathbb{R}^{D'\times D_m}$ and $\mathbf{b}_m\in\mathbb{R}^{D'\times 1}$ denote the activation function~\cite{LeakyReLU}, trainable weight matrix and bias vector, respectively. And $D'$ is the dimension of distilled feature vector $\bar{\mathbf{i}}_m$. 

Then, we introduce the neighbor routing mechanism~\cite{DisenGCN} into prototypical network, to approach the prototype \textit{w.r.t.} representation of user preference. 
Given a user, with the iterative routing operations, her/his representation is adjusted by jointly analyzing her/his similarities to its neighbors. 
To facilitate the description, we elaborate on the process in the single modality and do the same operations on the others. 

In the initial iteration, we define a trainable vector $\mathbf{u}_{(0)}$ to represent the preference of user $u\in\mathcal{U}$. 
And, we conduct the inner product between user preference and item features to calculate their similarity, formally,  
	\begin{equation}
		p_{u,i} = \frac{\exp(\bar{\mathbf{i}}^\mathsf{T}\mathbf{u}_{(0)})}{\sum_{j\in\mathcal{N}(u)}\exp(\bar{\mathbf{j}}^\mathsf{T}\mathbf{u}_{(0)})},
	\end{equation}
where $p_{u,i}$ denotes the similarity between $u$ and $i$. 
A higher value suggests that the content signal more informative to the user preference modeling. 
In addition, $\mathcal{N}(u)$ is used to represent the set of neighbors of node $u$ in the user-item graph.

Following this, we tune the representation of user preference in the metric space via combining the weighted sum of its neighbors' feature vectors. 
It is formulated as, 
	\begin{equation}
		\mathbf{u}_{(1)} =\mathbf{u}_{(0)}+\sum_{i\in\mathcal{N}(u)}p_{u,i}\bar{\mathbf{i}},
	\end{equation}
where $\mathbf{u}_{(1)}$ is the user representation after one iteration operation. 
Moreover, we normalize it to avoid its scale of increasing with iterative operations.

With the iteration $t=2,\dots,T$, based on the output of previous iteration, the user representation is adjusted towards the prototype of her/his preference, which is recursively formulated as: 
	\begin{equation}
    \begin{cases}
		\mathbf{u}_{(t)} =\mathbf{u}_{(t-1)}+\sum_{i\in\mathcal{N}(u)}p_{u,i}\bar{\mathbf{i}},\\[3mm]
		p_{u,i} = \frac{\exp(\bar{\mathbf{i}}^\mathsf{T}\mathbf{u}_{(t-1)})}{\sum_{j\in\mathcal{N}(u)}\exp(\bar{\mathbf{j}}^\mathsf{T}\mathbf{u}_{(t-1)})}.
	\end{cases}
	\end{equation}
Finally, it outputs user preference to the item content, as $\bar{\mathbf{u}}=\mathbf{u}_{(T)}$. 
In what follows, we use $\bar{\mathbf{u}}$ to denote the user preference to the content information. 
\vspace{3pt}\par\noindent\textbf{Pruning Operations.} 
To identify noisy edges, we score the affinity between user preference and item content to measure the confidence of the corresponding edge being true-positive interaction in each modality. 
Then, we integrate the scores of each edge in multiple modalities to yield the weight and assign it to the edge, which implements the pruning operations in a soft manner. 

For each modality, with the obtained user preference and distilled item features, we calculate the relative distances between them in two directions. 
It is formulated as,
	\begin{equation}
		\begin{cases}
		\bar{s}^m_{u\leftarrow i} = \frac{\exp(\bar{\mathbf{u}}_m^\mathsf{T}\bar{\mathbf{i}}_m)}{\sum_{j\in\mathcal{N}(u)}\exp(\bar{\mathbf{u}}_m^\mathsf{T}\bar{\mathbf{j}}_m)},\\[5mm]
	    \bar{s}^m_{i\leftarrow u} = \frac{\exp(\bar{\mathbf{i}}_m^\mathsf{T}\bar{\mathbf{u}}_m)}{\sum_{v\in\mathcal{N}(i)}\exp(\bar{\mathbf{i}}_m^\mathsf{T}\bar{\mathbf{v}}_m)},
		\end{cases}
	\end{equation}
where $\bar{s}^m_{u\leftarrow i}$ and $\bar{s}^m_{i\leftarrow u}$ are the scores reflecting the affinities between $\bar{\mathbf{u}}_m$ and $\bar{\mathbf{i}}_m$ in $m$-th modality. 

To integrate the multimodal scores, we define a base vector for each user or item, as follows:
\begin{equation}
    \bm{\rho} = [\ \rho^v,\ \rho^a,\ \rho^t]\ ,
\end{equation}
where $\bm{\rho}$ denotes the base vector. 
Elements of the user's base vector are used to measure her/his relative preferences to the different modalities. 
For the item's base vector, each element represents the importance of content signal in the corresponding modality to the item representation.

Incorporated base vectors, the weights for the edges are computed by fusing the multimodal scores, as
	\begin{equation}
		\begin{cases}
		s_{u\leftarrow i} = \max(\rho^v_u \bar{s}^{\;v}_{u\leftarrow i}\ ,\,\rho^a_u \bar{s}^{\;a}_{u\leftarrow i}\ ,\,\rho^t_u \bar{s}^{\;t}_{u\leftarrow i}),\\[3mm]
		s_{i\leftarrow u} = \max(\rho^v_i \bar{s}^{\;v}_{i\leftarrow u}\ ,\,\rho^a_i \bar{s}^{\;a}_{i\leftarrow u}\ ,\,\rho^t_i \bar{s}^{\;t}_{i\leftarrow u}).
		\end{cases}
	\end{equation}
where $\max(\cdot)$ denotes maximization operation selecting the max value. 
Besides, the combination operation is also able to implement in different forms, such as mean and maximization operations without base values.

In summary, with the base vector and obtained affinity scores, we achieve the weight for each edge to softly prune the noisy edge. 

\subsubsection{Graph Convolutional Layer}
Following the mainstream of GCN-based models~\cite{GCMC,NGCF},  we treat the graph convolutional operations as the message passing and aggregation. 
Using the graph convolutional operations, we could model the collaborative signal conveyed by user-item interaction graph. 
Further, by running the stacked graph convolutional layers, the high-order connectivity information is captured and aggregated. 
Towards the implicit feedback, the obtained weights for the edges are used to control the passed message. In particular, it corrupts the propagation of noise signal from false-positive interaction. 

Formally, in the $l$-th layer, the message passing and aggregation could be formulated as,
\begin{equation}
		\begin{cases}
			\mathbf{e}^{(l)}_u = \sum_{i\in\mathcal{N}(u)}s_{u\leftarrow i}\mathbf{e}^{(l-1)}_i,\\[3mm]
			\mathbf{e}^{(l)}_i = \sum_{u\in\mathcal{N}(i)}s_{i\leftarrow u}\mathbf{e}^{(l-1)}_u.
		\end{cases}
	\end{equation}
where $\mathbf{e}\in\mathbb{R}^{D\times 1}$ denotes the corresponding ID embedding vector. 
With this operation, we collect the collaborative signal from $l$-hop neighbors. 

Stacking L layers, we obtain the embedding at each layer and integrate them:
	\begin{equation}
		\mathbf{e}_u = \sum^{L}_{l=0}\mathbf{e}^{(l)}_u \, ,\ \ \mathbf{e}_i = \sum^{L}_{l=0}\mathbf{e}^{(l)}_i.
	\end{equation}
Whereinto, $\mathbf{e}_u^{(0)}$ and $\mathbf{e}_i^{(0)}$ denote the initial ID embeddings from the embedding matrix $\mathbf{E}$, respectively. 
The enriched embeddings~(\textit{i.e.} $\mathbf{e}_u$ and $\mathbf{e}_i$) are constituted by combing the embeddings from $0$-th layer to $L$-th layer. 
It encodes and injects the high-order connectivity information into the embedding of each node to enhance the representativeness. 
\subsubsection{Prediction Layer}
To gain the representation of each user or item, we follow the idea that users have varying preferences in different modalities~\cite{MMGCN}. 
Specifically, we concatenate the multimodal features and the enriched ID embedding as a whole vector, formally,
	\begin{equation}
		\begin{cases}
			\mathbf{e}^*_u = \mathbf{e}_u\,\lVert\,\bar{\mathbf{u}}_v\,\lVert\,\bar{\mathbf{u}}_a\,\lVert\,\bar{\mathbf{u}}_t,\\[3mm]
			\mathbf{e}^*_i = \mathbf{e}_i\;\lVert\;\bar{\mathbf{i}}_{\,v}\;\lVert\;\bar{\mathbf{i}}_{\,a}\;\lVert\;\bar{\mathbf{i}}_{\,t},
		\end{cases}
	\end{equation}
where the symbol $||$ means the concatenation operation. 

Beyond the collaborative signals, the representation contains the user preference to the item content, which contributes to inferring the interaction between users and items. 

Finally, we conduct the inner product between user and item representations, as 
\begin{equation}
    y_{u,i} = {e^*_u}\mathsf{T}\  {e^*_i},
\end{equation}
where the output $y_{u,i}$ is used to estimate the user's preference towards the target item. 
A higher score suggests that the user prefers the item more and vice versa.  
\subsection{Optimization}
To learn the parameters of the proposed model, we adopt Bayesian Personalized Ranking~(BPR)~\cite{BPR} to conduct the pair-wise ranking. 
As such, we construct a triplet of one user $u$, one observed item $i$, and one unobserved item $j$, formally as, 
\begin{equation}
    \mathcal{T} = \{(u, i, j)\ | \mathbf{A}_{u,i}=1,\ \mathbf{A}_{u, j}=0 \},
\end{equation}
where $\mathcal{T}$ is a triplet set for training. 

Therefore, the objective function can be defined as, 
\begin{equation}
    \mathcal{L} = \sum_{(u, i, j)\in\mathcal{T}}{-\ln \phi(y_{u,i} - y_{u,j})}+\lambda\left\|\theta\right\|_2,
\end{equation}
where $\phi(\cdot)$, $\lambda$, and $\theta$ represent the $sigmoid$ function, regularization weight and parameters of models, respectively. 
\begin{table}
\small
  \centering
  \caption{Summary of the datasets. The dimensions of visual, acoustic, and textual modalities are denoted by V, A, and T, respectively.}
      \vspace{-5pt}
  \label{table_1}
  \setlength{\tabcolsep}{1.1mm}
  \begin{tabular}{c|c|c|c|c|c|c|c}
    \specialrule{0.1mm}{0pt}{1pt}
    Dataset& \#Interactions & \#Items & \#Users & Sparsity & V & A & T\\
    \specialrule{0.1mm}{1pt}{1pt}
    \specialrule{0.1mm}{1pt}{1pt}
    Movielens&1,239,508&5,986&55,485&99.63\%&2,048&128&100\\
    \specialrule{0.0mm}{0pt}{1pt}
    Tiktok&726,065&76,085&36,656&99.97\%&128&128&128\\
    \specialrule{0.0mm}{0pt}{1pt}
    Kwai&298,492&86,483&7,010&99.95\%&2,048&-&-\\
    % \hline
    \hline
  \end{tabular}
\end{table}
\begin{table*}
  \centering
  \renewcommand\arraystretch{1}
  \setlength{\tabcolsep}{3mm}
  \caption{Performance comparison between our model and the baselines over the three datasets.}
  \vspace{-5pt}
  \label{table_2}
  \begin{tabular}{c|ccc|ccc|ccc}

    \hline
    \multirow{2}{*}{Model}&\multicolumn{3}{c|}{Movielens}&\multicolumn{3}{c|}{Tiktok}&\multicolumn{3}{c}{Kwai}\\
    &Precision&Recall&NDCG&Precision&Recall&NDCG&Precision&Recall&NDCG\\
    \hline
    \hline
    GraphSAGE&0.0496&0.1984&0.2136&0.0128&0.0631&0.0606&0.008&0.0286&0.0467\\
    MMGCN&\textbf{0.0581}&\textbf{0.2345}&\textbf{0.2517}&0.0144&0.0808&0.0674&0.0120&0.0398&0.0681\\
    \hline
    NGCF&0.0547&0.2196&0.2342&0.0135&0.0780&0.0661&0.0118&0.0402&0.0699\\
    DisenGCN&0.0555&0.2222&0.2401&0.0145&0.0760&0.0639&0.0127&0.0403&0.0683\\
    GAT&0.0569&0.2307&0.2434&\textbf{0.0166}&\textbf{0.0891}&\textbf{0.0802}&\textbf{0.0151}&\textbf{0.0441}&\textbf{0.0744}\\
    \hline
    Ours&\textbf{0.0639}*&\textbf{0.2569}*&\textbf{0.2754}*&\textbf{0.0195}*&\textbf{0.1048}*&\textbf{0.0938}*&\textbf{0.0168}*&\textbf{0.0492}*&\textbf{0.0864}*\\
    \specialrule{0.1mm}{1pt}{1pt}  
    \specialrule{0.1mm}{1pt}{1pt}  
    \%Improv.&\textbf{9.98\%}&\textbf{9.55\%}&\textbf{9.42\%}&\textbf{17.47\%}&\textbf{17.62\%}&\textbf{16.96\%}&\textbf{11.26\%}&\textbf{11.56\%}&\textbf{15.66\%}\\
    \specialrule{0.1mm}{1pt}{1.5pt}  
  \end{tabular}\vspace{-5pt}
\end{table*}
\section{EXPERIMENTS}
Through conducting extensive experiments on three public datasets, we evaluate our proposed model and answer the following research questions:
\begin{itemize}
    \item \textbf{RQ1} How does our proposed model perform compared with state-of-the-art GCN-based recommendation models?
    \item \textbf{RQ2} How does each design (\textit{i.e.} prototypical network and  pruning operations) affect the performance of our model?
    \item \textbf{RQ3} How does the representation benefit from the refined graph? 
\end{itemize}

Before answering the above three questions, we describe the datasets, evaluation protocols, baselines, and parameter settings in the experiments.

\subsection{Experiments Settings}

\subsubsection{Dataset}
As the micro-video contains rich multimedia information — frames, sound tracks,
and descriptions~\cite{LM1,LM2,Nie2017Enhancing,NMCL}, we performed the micro-video personalized recommendation to evaluate our proposed method~\cite{JH}. 
Following MMGCN, we conducted extensive experiments on three publicly accessible datasets, including Movielens, Tiktok, and Kwai. 
The statistics of datasets are summarized in Table~\ref{table_1}. 
\begin{itemize}
    \item \textbf{Movielens.}
    The dataset is widely used in the personalized recommendation\footnote{https://movielens.org/.}. 
    To evaluate the multimedia recommendation, researchers extracted keyframes and soundtracks from the video trailers, as well as collected the video descriptions~\cite{MMGCN}. 
    With some pre-trained deep learning models~\cite{ResNet,VGG,S2V}, the visual, acoustic, and textual features are captured from the keyframes, audio tracks, and descriptions, respectively. 
    % It contains $1,239,508$ of ratings from $55,485$ users on $5,986$ movies. 
    In our experiments, we treat all ratings as the implicit feedback of the corresponding user and item pairs.  
 
    \item \textbf{Tiktok.} 
    This dataset is released by Tiktok\footnote{https://www.tiktok.com/.} which is a popular micro-video sharing platform. 
    % It consists of $36,656$ users, $76,085$ micro-videos, and $726,065$ users' historical browsing records. 
    Beyond the interaction information, the visual, acoustic, and textual features are extracted from the micro-videos and provided.   
    
    \item \textbf{Kwai.}
    As a micro-video service provider, Kwai\footnote{https://www.kwai.com/.} released a large-scale micro-video dataset.
    The dataset contains users, micro-videos, and the users' behavior records with the timestamps. 
    % ~(\textit{i.e.} clicks, likes, and follows). 
    To evaluate the proposed method from implicit feedback, we collected some click records associated with the corresponding users and micro-videos in a certain period.  
    % Finally, we obtained $7,010$ users, $86,483$ items, and $298,492$ interactions for our experiments. 
    Different from the above datasets, the audio and textural features are not given. 
\end{itemize}
For each dataset, we used the ratio $8:1:1$ to randomly split the historical interactions of each user and constituted the training set, validation set, and testing.  
For the training set, we conducted a negative sampling strategy to create the triples for parameter optimization. 
The validation set and testing set are used to tune the hyper-parameters and evaluate the performance in the experiments, respectively.  
\subsubsection{Evaluation Protocols}
For each user in the validation and the testing sets, we treated all micro-videos she/he did not consume before as the negative samples. 
During the validation and testing phases, we used the trained model to score the interactions of user and micro-video pairs and ranked them in a descending order. 
Moreover, following the widely-used evaluation metrics~\cite{SML,KGPolicy,DGCF,MAML}, we adopted precision@K, recall@K, and Normalized Discounted Cumulative Gain~(NDCG@K) to evaluate the performance of  methods. 
By default, we set $K=10$ and reported the average values of the three metrics for all users in the test set. 

\subsubsection{Baselines}
To evaluate the boosting of our proposed model towards the GCN-based recommendation, we compared it with several state-of-the-art GCN-based models for recommendation with implicit feedback. 
We briefly divided them into two groups: message-nonadaptive methods~(\textit{i.e.} GraphSAGE and MMGCN) and message-adaptive methods~(\textit{i.e.} NGCF,  GAT, and DisenGCN).
For a fair comparison, we employed two graph convolutional layers in our proposed model and baselines.

\begin{itemize}
    \item \textbf{GraphSAGE}~\cite{GraphSAGE}
    We applied GraphSAGE on the user-item graph from implicit feedback to predict the interaction between user and item. 
    With the trainable aggregation functions, the model is able to pass the message along the graph structure and collect them to update the representation of each node. 
    \item \textbf{MMGCN}~\cite{MMGCN}
    The model learns the model-specific user preference to the content information via the directly information interchange between user and item in each mormaitidality. 
    Based on the assumption that the user prefers the observed item over the unobserved one, it is trained from implicit feedback. 
    \item \textbf{NGCF}~\cite{NGCF}
    As the state-of-the-arts GCN-based model of personalized recommendation, NGCF explicitly models and injects the collaborative signal into user and item embeddings. 
    It measures the distance between user and item embeddings to control the passed message.   
    \item \textbf{DisenGCN}~\cite{DisenGCN}
    The method could disentangle the representation of each node into several factors. 
    By dynamically identifying the correlation of factor between nodes, it assigns the different weights to edges and aggregates the message to improve the representation of each node.
    \item \textbf{GAT}~\cite{GAT}
    This method is able to automatically learn and specify different weights to the neighbors of each node. 
    With the learned weights, it denoises the information from the neighbors to improve the personalized recommendation. 
\end{itemize}
\begin{figure}
    \centering
    \subfigure[Recall@10 on Movielens]{
      \includegraphics[width=0.225\textwidth]{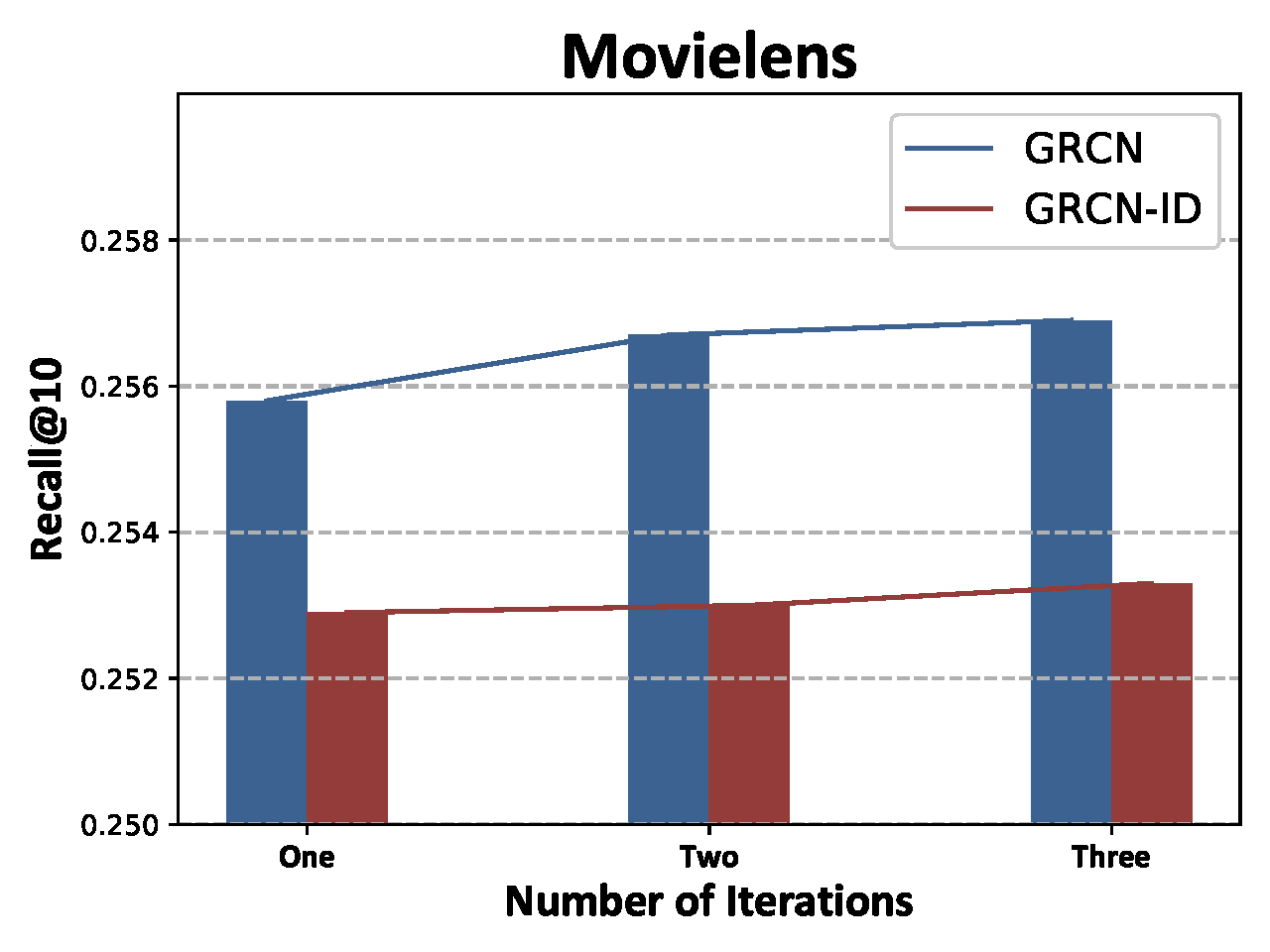}
      \label{fig_visualize_1_1}
    }
    \subfigure[NDCG@10 on Movielens]{
      \includegraphics[width=0.225\textwidth]{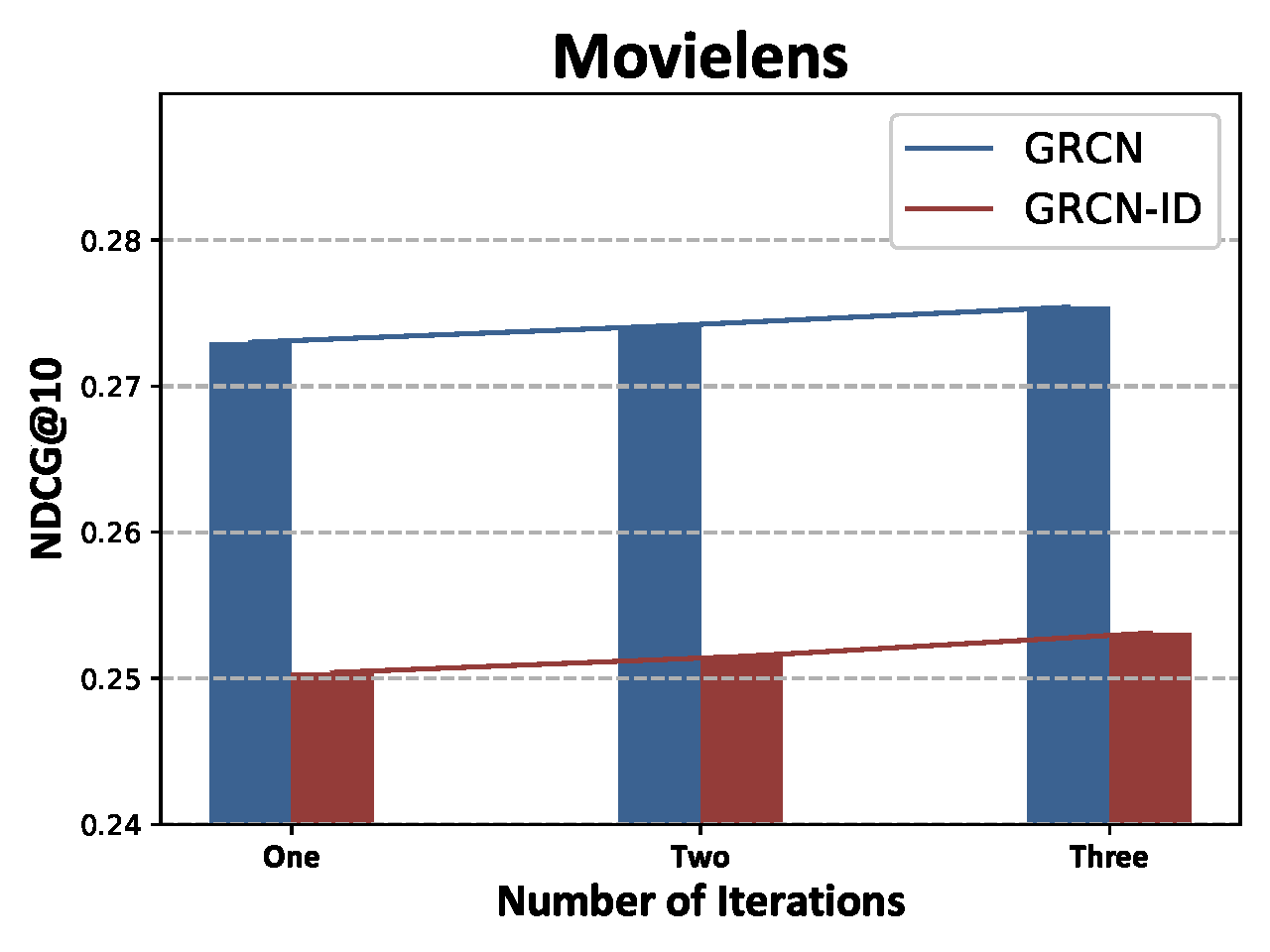}
      \label{fig_visualize_2_1}
    }
    \subfigure[Recall@10 on Tiktok]{
      \includegraphics[width=0.225\textwidth]{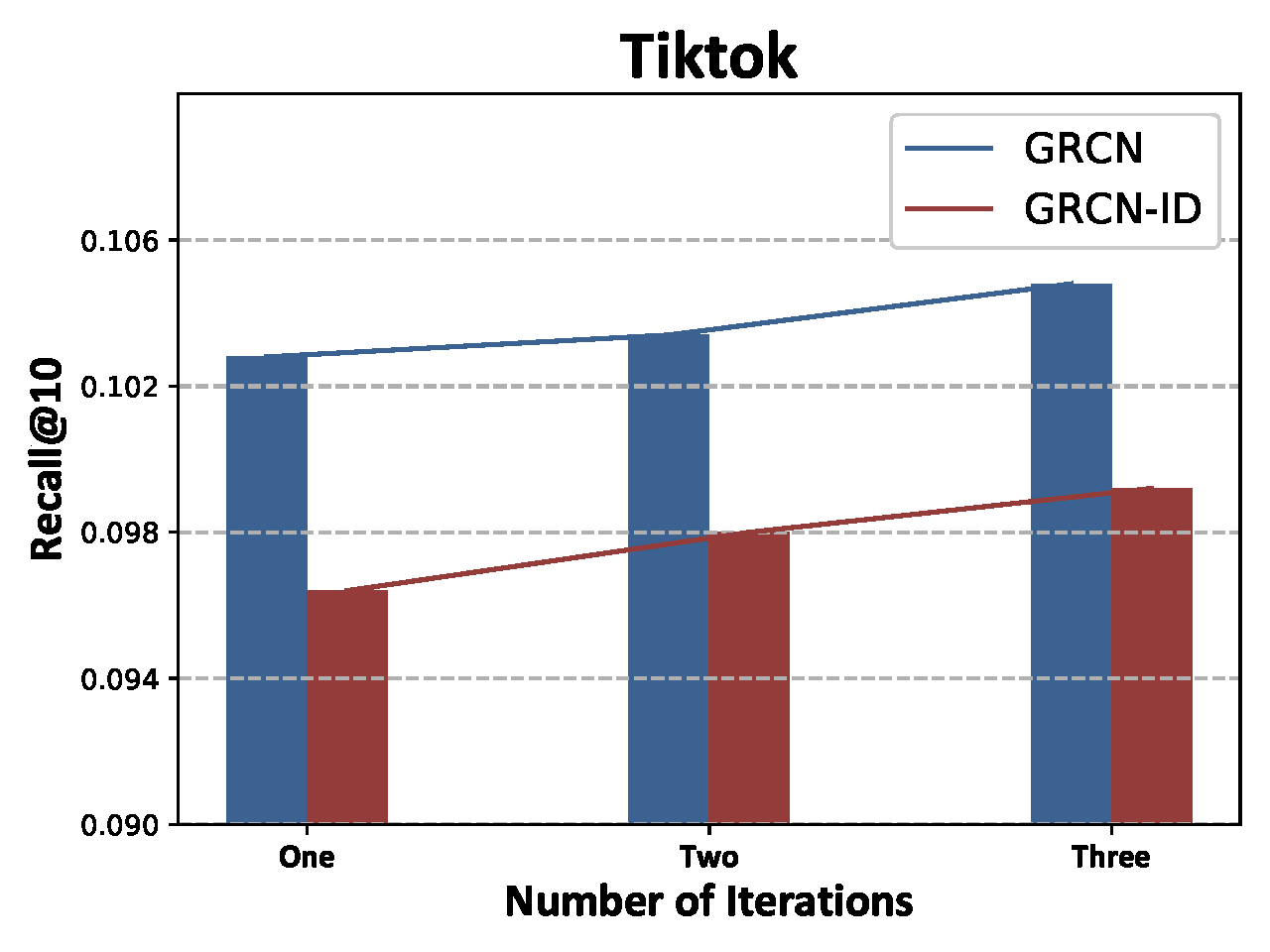}
      \label{fig_visualize_1_2}
    }
    \subfigure[NDCG@10 on Tiktok]{
      \includegraphics[width=0.225\textwidth]{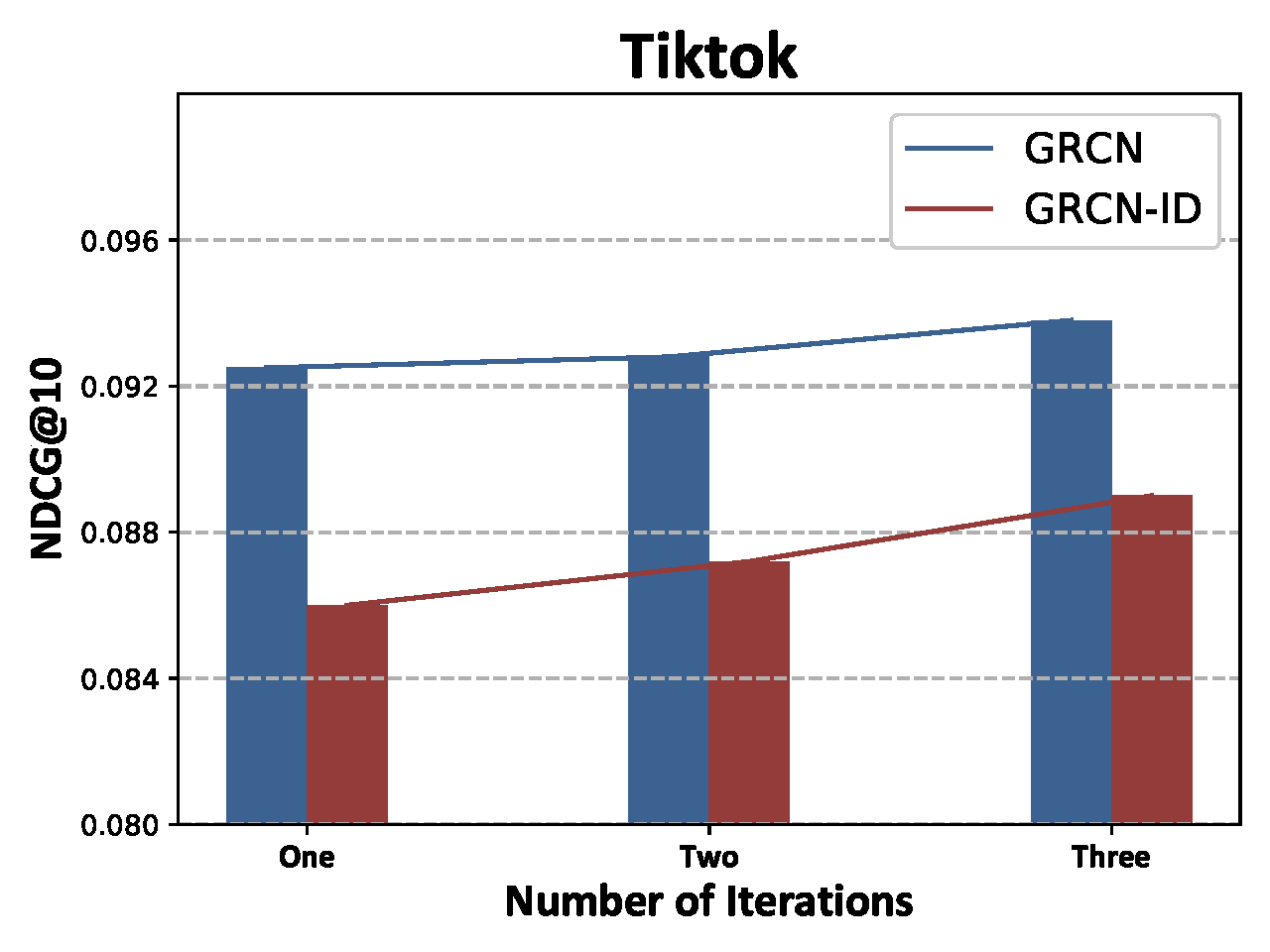}
      \label{fig_visualize_2_2}
    }
    \subfigure[Recall@10 on Kwai]{
      \includegraphics[width=0.225\textwidth]{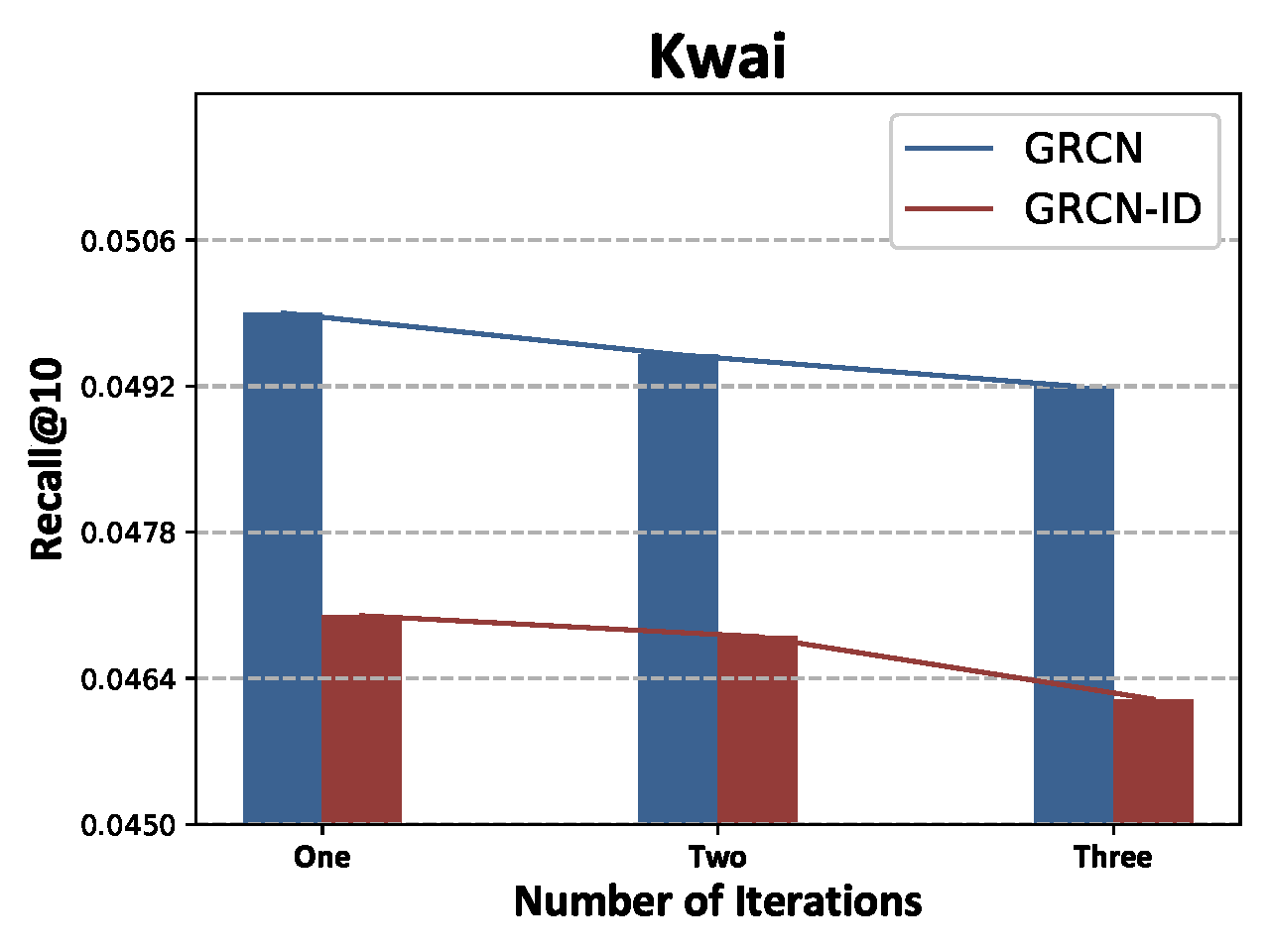}
      \label{fig_visualize_1_3}
    }
    \subfigure[NDCG@10 on Kwai]{
      \includegraphics[width=0.225\textwidth]{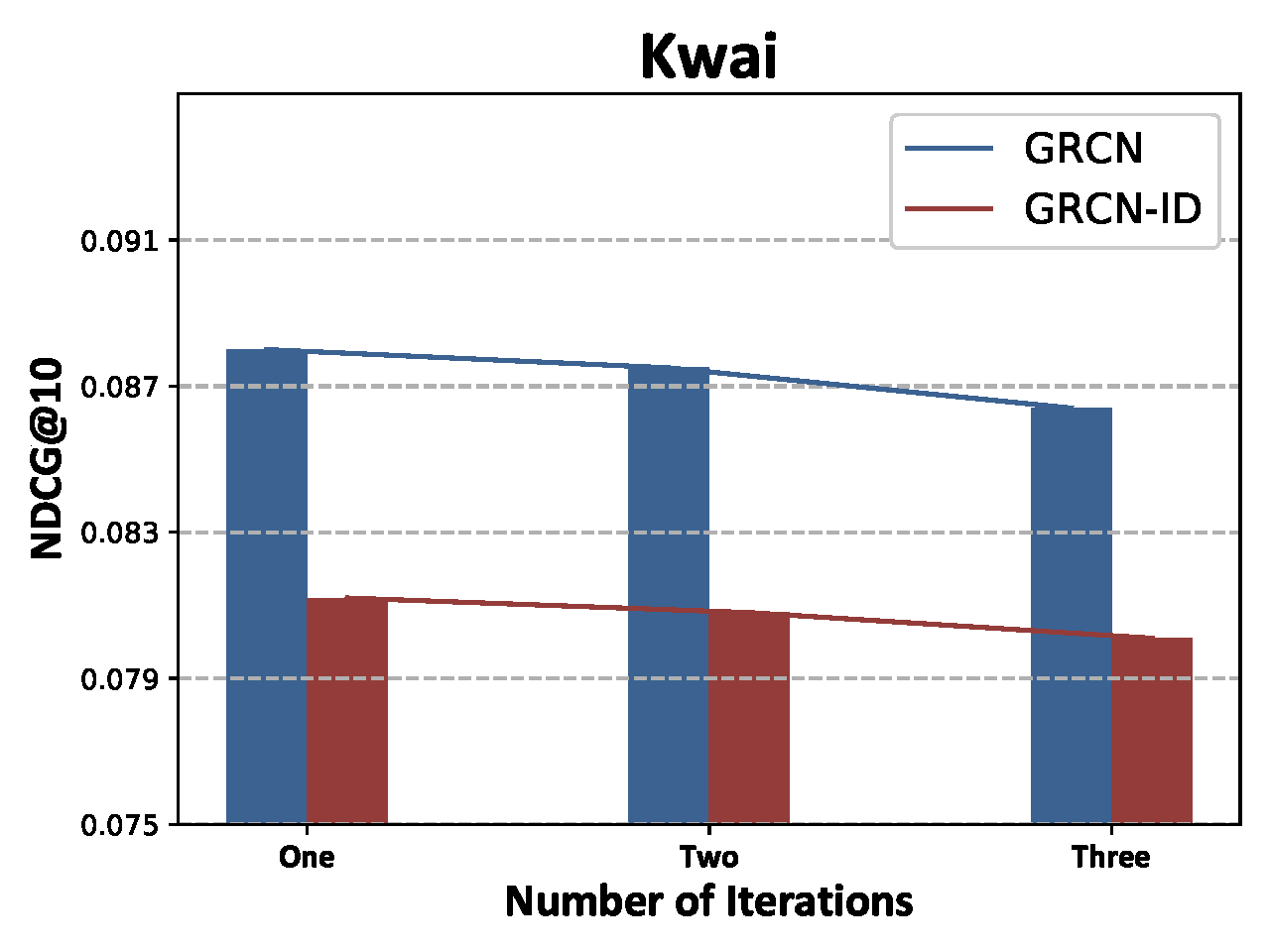}
      \label{fig_visualize_2_3}
    }
    \vspace{-10pt}
    \caption{Performance in terms of Recall@10 and NDCG@10 w.r.t. different numbers of iterations performing prototypical network on Movielens, Tiktok, and Kwai.}
    \label{fig_3}
    \vspace{-5pt}
 \end{figure} 
\subsubsection{Parameter Settings}
The Pytorch\footnote{https://pytorch.org/.} and torch-geometric packages\footnote{https://pytorch-geometric.readthedocs.io/.} are utilized to implement our proposed model. 
We used the Xavier~\cite{Xavier} and Adam~\cite{Adam} methods to initialize and optimize parameters, respectively. 
In addition, the learning rate is searched in $\{0.0001, 0.001, 0.01, 0.1, 1\}$ and regularization weight is tuned in $\{0.00001, 0.0001, 0.001, 0.01, 0.1\}$. 
Besides, we stopped the training if recall@10 on the validation data does not increase for 20 successive epochs. 
For the baselines, we followed the designs in their articles to achieve the best performance. 
Further, we did the same options and fixed the dimension of the ID embedding vector to 64 for all models to ensure a fair comparison.  
\subsection{Performance Comparison~(RQ1)}
Table~\ref{table_2} reports the performance comparison results. We have the following observations:
\begin{itemize}
    \item Without any doubt, our proposed model consistently achieves the best performance on three datasets. 
    In particular, the method improves over the strangest baselines in terms of Recall@10 by $9.55\%$, $17.62\%$, and $11.56\%$ in Movielens, Tiktok, and Kwai, respectively. 
    With the similar graph convolutional operations, the improvements could be attributed to the graph refining. 
    It verifies our suggestion that identifying and pruning the noisy edges in the interaction graph benefits the GCN-based recommendation model. 
    \item Comparing with GraphSAGE, message-adaptive methods, including NGCF, GAT, DisenGCN, and GRCN, yield better results. It implies that the message during the passing and aggregation process conveys some meaningless or harmful signal for the node representation. In other words, the local structure information in the graph constructed by implicit feedback cannot directly reflect the user preference pattern. 
    \item It is worthwhile pointing that MMGCN slightly outperforms other baselines in several cases. 
    We believe one possible reason is that MMGCN sufficiently leverages the multimodal information extracted from items to represent the user preference. 
    Although the method ignores the perturbations in the structure of graph caused by implicit feedback, it implicitly distills the information during its cross-modalities combination layers. 
    The results on Kwai, which only contains the visual modality, could be used to verify this point. 
    \item Obviously, adaptively adjusting the passed message in the graph convolutional operations contributes to the user and item modeling in recommendation with implicit feedback. 
    However, compared with GRCN, other message-adaptive models are suboptimal. 
    We argue the message yielded by these models is corrupted by the graph structure. 
    Specifically, the uncertain message totally depending on initial nodes' ID embeddings is alternately propagated from one node to the other over each edge. 
    Rather, our proposed method measures the affinity between user preference and item content to control the message passing before iteratively conducting the graph convolution layers.  
    
\end{itemize}
\subsection{Ablation Study~(RQ2)}
In this section, we evaluated the designs of our proposed model, especially the graph refining layer. 
The graph refining layer is devised to model the user preference from content information and refine the interaction graph for optimizing the GCN-based recommendation method. 
As such, we conducted experiments to evaluate two components in the refining layer, including the prototypical network and pruning operations. 
\subsubsection{Effects of Prototypical Network}
To evaluate the effect of prototypical network on user preference learning, we performed the experiments under different numbers of routing iterations. 
Meanwhile, we compared the proposed model with the variant, marked as GRCN-ID, which discarded the multimodal user preference and item features in the prediction layer and directly predicted the interaction between user and item with only their ID embeddings. 
As illustrated in Fig~\ref{fig_3}, we observe that:
\begin{itemize}
    \item With the increasing of iterations, the values in terms of Recall@10 and NDCG@10 are varying. 
    It is indicated that the performance of our proposed method is affected by user preference modeling. 
    \item On Movielens and Tiktok, the values are increasing when we iteratively perform the prototypical network.  
    We believe that the user representation is tuned towards her/his preference to the content information, which benefits the correlation computing and graph refining. 
    \item For Kwai, the curves of results \textit{w.r.t} Recall@10 and NDCG@10 tend to decline with the iterative operations. 
    We suggest the phenomenon is caused by the smooth representation of users. 
    In this dataset, the number of average interaction records of a user is much more than the others. 
    Hence, the iterative operations make the representation smooth, which is hard to capture the discriminative features for users. 
    \item Obviously, GRCN outperforms GRCN-ID, which benefits from the user~(item) representation enriched by corresponding user preference~(item features). 
    Although GRCN-ID is suboptimal on three datasets, it is still better than the abovementioned GCN-based baselines. 
    Jointly analyzing the performance of baselines shown in Tabel~\ref{table_2}, it qualitatively verifies the refined graph structure contributes to the GCN-based recommendation model. 
\end{itemize}
 \begin{table}
  \centering
  \renewcommand\arraystretch{1}
  \setlength{\tabcolsep}{0.7mm}
  \caption{Effect of pruning operations on Movielens and Tiktok. (Visual, Acoustic, and Textual denote running GRCN on the visual, acoustic, and textual modality, respectively.)}
  \vspace{-3mm}
  \label{table_3}
  \begin{tabular}{c|ccc|ccc}%s|ccc}
    \hline
    \multirow{2}{*}{Model}&\multicolumn{3}{c|}{Movielens}&\multicolumn{3}{c}{Tiktok}\\%&\multicolumn{3}{c}{Kwai}\\
    &\small{Precision}&\small{Recall}&\small{NDCG}&\small{Precision}&\small{Recall}&\small{NDCG}\\%&Precision&Recall&NDCG\\
    \hline
    \specialrule{0.1mm}{1pt}{1.5pt}
    Visual&0.0633&0.2545&0.2714&0.0175&0.0906&0.0822\\
    \specialrule{0.0mm}{0pt}{1pt}
    Acoustic&0.0621&0.2540&0.2701&0.0144&0.0788&0.0694\\
    \specialrule{0.0mm}{0pt}{1pt}
    Textual&0.0611&0.2531&0.2648&0.0142&0.0770&0.0675\\
    \hline
    \specialrule{0.0mm}{0pt}{1pt}
    $GRCN_{max}$&0.0621&0.2542&0.2701&0.0175&0.0941&0.0838\\
    \specialrule{0.0mm}{0pt}{1pt}%\uppercase\expandafter{\romannumeral2}
    $GRCN_{mean}$&0.0617&0.2477&0.2660&0.0159&0.0854&0.0751\\    
    \specialrule{0.0mm}{0pt}{1pt}%\uppercase\expandafter{\romannumeral2}
    $GRCN_{hard}$&0.0639&0.2547&0.2750&0.0180&0.0962&0.0868\\ 
    \specialrule{0.1mm}{1pt}{1.5pt}        
    \specialrule{0.1mm}{0pt}{1pt}
    GRCN&\textbf{0.0639}&\textbf{0.2569}&\textbf{0.2754}&\textbf{0.0195}*&\textbf{0.1048}*&\textbf{0.0938}*\\%&\textbf{-}*&\textbf{-}*&\textbf{-}*\\
    \hline
  \end{tabular}
    \vspace{-5mm}
\end{table}
 \subsubsection{Effects of Pruning Operations.}
 To explore the pruning operations, we compared the performance of our proposed model with three different implementations. 
 Specifically, we adopted the mean and maximization operations without the base value, which are named as $\text{GRCN}_{mean}$ and $\text{GRCN}_{max}$, to fuse the multimodal affinity scores, respectively.  
 Besides, we compared our model with the hard pruning strategy which is labeled as $\text{GRCN}_{hard}$ and implemented with ReLU function~\cite{ReLU} to completely interrupt the message passed from the false-positive interaction. 
 In addition, we conducted GRCN in each modality as a comparison.
 From the results in Table~\ref{table_3}, we have following observations:
 \begin{itemize}
    \item In most cases, we observe that the results of three implements are better than that of models within the single modality. 
    It demonstrates that incorporating the information from multiple modalities facilitates the pruning operations, since users have various opinions about the different modalities of micro-videos. 
    \item Observing the results of $\text{GRCN}_{mean}$ and $\text{GRCN}_{max}$, we find that the later is superior to the former. 
    Especially, the results of $\text{GRCN}_{mean}$ significantly underperform the model which merely considers the visual modality. 
    It probably implies that the maximization operation is consistent with the relationship among different modalities and able to model it. 
     \item 
     Both our proposed model and its variant $\text{GRCN}_{hard}$ outperform the other two implements without the base value. 
     It shows that incorporating the base value is capable of boosting performance, which justifies our purpose regarding the base values. 
	\item As expected, compared with other variants, our proposed model yields the best results. 
	Beyond the multimodal information and base value incorporating, it also makes sense because of the soft pruning operation. 
	Different from the hard pruning, pruning operation in a soft manner not only weakens the noise caused by false-positive interaction but enhances the message from true-positive ones. 
	It contributes to refine the graph structure for graph convolutional operations. 
 \end{itemize}

\begin{figure}
    \centering
    \subfigure[GAT]{
      \includegraphics[width=0.22\textwidth]{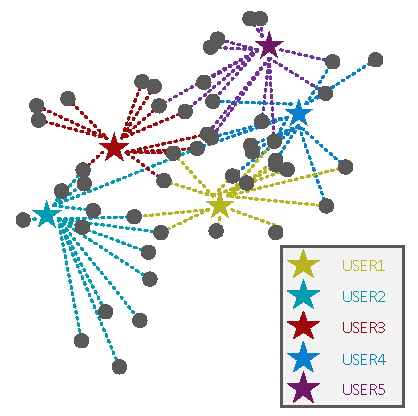}
      \label{fig_visualize_4_1}
    }
    \subfigure[GRCN]{
      \includegraphics[width=0.23\textwidth]{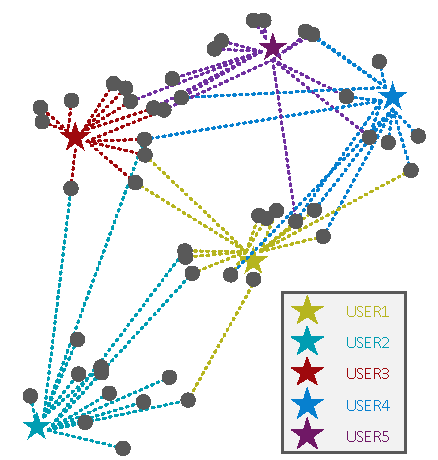}
      \label{fig_visualize_4_2}
    }
    \caption{Visualization of the learned t-SNE transformed representations derived from GAT and GRCN. The star marks denote the user randomly selected from Tiktok. The link between star and circle means the interaction between them.}
    \label{fig_4}
    \vspace{-5pt}
 \end{figure} 
\subsection{Visualization}
In this section, we provided several users randomly selected from Tiktok associated with their consumed items. 
Using the t-Distributed Stochastic Neighbor Embedding~(t-SNE) in 2-dimension, we visualized their ID embeddings, which are learned from GAT and GRCN, respectively. 
Besides, we used same color to denote edges from a user node to the nodes of items she/he interacted with. As illustrated in Figure~\ref{fig_4}, we have two key observations:
\begin{itemize}
    \item From the distribution of nodes in two figures, we find that the nodes representing the items exhibit discernible clustering around the user nodes in Figure~\ref{fig_visualize_4_2}. 
    It means that GRCN discriminately represent the users, although there are several items consumed by the same users. 
    We attribute this to the graph refining operations assigning different weights for edges according to affinities between corresponding user preference and item content, which eliminates the noises from false-positive interactions. 
    \item By observing Figure~\ref{fig_visualize_4_1}, it is shown that the distribution of users is closer. 
    Theoretically, GAT should be able to model the correlation between user and item and distinguish them with the help of attention mechanism. 
    Unexpectedly, it pulls the nodes of users who consumed the same items closer. 
    The reason I suggest is that GAT heavily affected by the initialization of ID embeddings, which is consistent with the finding~\cite{AttUnder}. 
    On the contrary, our proposed method is stable and efficient to capture the correlation between users and items. 
    
\end{itemize}
\section{Related Work}
In this section, we review the existing work related to our research, including the recommendation with implicit feedback and GCN-based personalized recommendation.
\subsection{Recommendation with Implicit Feedback}
Comparing with the work focusing on explicit feedback~\cite{Ex1,Ex2,Yu}, profiling the user from implicit feedback is more practical and challenge. 
Therefore, researchers shift to explore the user-item interaction from implicit feedback data. 

To address the challenge of implicit feedback, the core is how to distinguish the negative instances from the positive ones. 
As such, Hu~\textit{et al.}~\cite{WRMF} treated the all user's unobserved item as negative instances and indicated the numerical value of implicit feedback as confidence. 
Besides, Rendle~\textit{et al.}~\cite{BPR} proposed Bayesian Personalized Ranking~(BPR) method, which sampled negative instances from the user's unobserved items to construct the triple of <user, positive item, negative item> for pair-wise ranking. 
Comparing with the method assigning a uniform weight to each user's unobserved item, He~\textit{et al.}~\cite{FastMF} proposed to weight them based on item popularity and designed a model to efficiently optimize with variably-weighted item. 
Recently, Yang~\textit{et al.}~\cite{HOPrec} treated the items belonging to user's high-order neighbors as positive instances and others as negative ones, which enriches the training set to optimize the parameters of proposed graph and matrix factorization~(MF) combination model. 

In terms of the multimedia personal recommendation, He~\textit{et al.}~\cite{VBPR} extended BPR method and proposed Visual Bayesian Personalized Ranking~(VBPR), which incorporated the visual information to improve the performance. 
Following BPR method, they used all user's unobserved items as negative instances and performed the pair-wise ranking operation. 
Beyond exploring the positive and negative instances from implicit feedback, Chen~\textit{et al.}~\cite{ACF} designed Attentive Collaborative Filtering~(ACF) model to capture item- and component-level implicit feedback in the multimedia recommendation. 

Different from the existing studies, we propose to model the user preference to the item content and measure the similarity between them to discover the false-positive feedback from the historical records.  
\subsection{GCN-based Personalized Recommendation}
Due to the effectiveness in representation learning, graph convolutional network are widely exploited to model the interactions between users and items for personalized recommendation. 
For instance, Berg~\textit{et al.}~\cite{GCMC} formulated the recommendation task as a link prediction problem on graphs and utilized the graph convolutional operation to predict links between user and item. 
Based on differentiable message passing on the bipartite graph, they devised a graph auto-encoder framework. 
Nevertheless, the method is designed for the recommender system with explicit feedback data~(\textit{i.e.} ratings).  

Towards the implicit feedback, Ying~\textit{et al.}~\cite{PinSAGE} constructed a bipartite interaction graph according to users' browsing records and developed a large-scale recommendation engine for image recommendation. 
On the constructed graph, the method jointly conducts the graph convolutional operations and random walks to represent the users and items, which supercharges the efficiency on web-scale personalized recommendation.
Similarly, Wang~\textit{et al.}~\cite{NGCF} constructed the user-item graph, whose edges corresponded to implicit feedback. 
With their proposed neural graph collaborative filtering~(NGCF) method, the collaborative signal conveyed by the edges and high-order connectivity explicit modelled and injected into each user and item embedding. 
More recently, the GCN-based model has been introduced into multimedia recommendation in implicit feedback settings. 
Wei~\textit{et al.}~\cite{MMGCN} constructed the modal-specific bipartite graph with implicit data to model the user preference in multiple modalities. 
They developed a multimedia recommendation framework, dubbed multimodal graph convolutional network~(MMGCN), which represented the user preference in each modality with her/his directly and indirectly connected neighbors. 

However, these GCN-based recommendation models ignore the effect of implicit feedback. 
Moreover, with iteratively performing the graph convlutional operations, the disruption of node representation, which is caused by distorting of graph structure, becomes worse. 
Against this issue, we propose to refine the user-item graph to eliminate its effect in the graph convolutional operations.

\section{CONCLUSION AND FUTURE WORK}
In this paper, we propose to solve the problem of implicit feedback towards the GCN-based recommendation method. 
Therefore, we develop a novel model, named Structure-Refined Graph Convolutional Networks, which yields a refined user-item interaction graph for graph convolutional operations. 
It identifies the false-positive feedback and prunes the corresponding noisy edge in the interaction graph. Empirical results on three public benchmarks demonstrate the efficiency of our proposed model.

To the best of our knowledge, this work is the first attempt to explore the disadvantage of the GCN-based recommendation caused by implicit feedback. 
Despite the state-of-the-art performance our proposed model achieved, we believe there is a long distance to solve the implicit feedback problem thoroughly. 
We attribute the issues caused by implicit feedback to the gap between user preference and behaviors. Beyond the user preference, the motivation why people prefer some items (\textit{i.e.} user intent) is critical to estimate the user behaviors, but inefficiently explored. 
As such, in future work, we expect to study how to learn and leverage the user intents, in order to provide a high-quality personalized recommender system. 
\bibliographystyle{ACM-Reference-Format}
\balance
\newpage
\bibliography{reference}
\balance
% \bibliography{BIB/IEEEabrv,MulGCN} 

\end{document}